# Humanitarian Algorithms : A Codified Key Safety Switch Protocol for Lethal Autonomy [12]

Nyagudi Musandu[1]

(Independent Researcher, Nairobi, KENYA – EAST AFRICA)

**Abstract:** With the deployment of lethal autonomous weapons, there is the requirement that any such platform complies with the precepts of International Humanitarian Law. Humanitarian Algorithms[9: p. 9] ensure that lethal autonomous weapon systems perform military/security operations, within the confines of International Humanitarian Law. Unlike other existing techniques of regulating lethal autonomy this scheme advocates for an approach that enables Machine Learning. Lethal autonomous weapons must be equipped with appropriate fail-safe mechanisms that locks them if they malfunction.

## 0 Introduction

The codified key protocol for activating lethal autonomous weapons using humanitarian algorithms goes beyond the Arkin Ethical Governor approach that acts as a "bottleneck" to already generated lethal impulses. It allows for the initiating of lethal action by an autonomous weapon system only upon complete verification that any such targeting action complies with Laws of War, Rules of Engagement, Ethics of Owner and Worldview of Owner – these parameters are used to set the activation rule parameters while the target characteristics act as the weapons release keys.

This protocol goes further to advance a process for multi-robot collaboration amongst several (semi)autonomous entities, by way of transfer and sharing of tasks in a battle-space. There are existing architectures such as the Joint Architecture for Unmanned Systems[1] but is not specifically designed and optimized for lethal autonomous robots. Hopefully this new safety switching protocol for lethal autonomous robots shall offer some suggestions about how an autonomous weapon can be deactivated/neutralized in the event of a cessation of hostilities. Other thought-provoking propositions would be the ability of the system to determine difficult to discern human traits such as treachery or courage and to register, record or report the same.

---

1 **Brief author introduction:** Nyagudi Musandu (1973 -), male, Independent researcher, humanitarian informatics. Email : nyagudim@yahoo.com

# Humanitarian Algorithms : A Codified Key Safety Switch Protocol for Lethal Autonomy [12]

## 1 Rationale

Over fifty of the world's nations are today developing and/or using tactical robots, these range from bomb disposal robots, aerial surveillance drones and unmanned underwater vehicles. The classification of unmanned autonomous weapons could be extended to include some classes of naval mines and tactical missile systems. Given the ubiquitous nature of lethal autonomy, it is a phenomena that should be harnessed for the right reasons and no ban is expected anytime in the near future, as it offers a natural progression from many current remote controlled platforms.

Mizroch [2] documents Israel Air Force's nascent attempts at defining the realm of lethal autonomy. Probably influenced by the Arkin Ethical Governor, there is the terminology Ethical Algorithms in the publication. The perspective put forward in this paper is that a lethal autonomous weapon can only be ethical to the extent that its user/owner is ethical. Mies [3] suggests that robots can replace man in battle-space for reasons of cost reduction, quality and efficacy at implementation of international humanitarian law, without detailing how this can be achieved in the foreseeable future.

Even with the seemingly pervasive deployment of autonomous weapons technologies, there are still those opposed to their full-scale adoption by military forces, some of whom are calling for a world-wide ban. Sharkey [4] strongly sets out the opinion that robots are not capable of distinguishing between combatants and non-combatants, a position that is not true to the extent that there are a wide range of machine learning methods. This paper suggests some configurations that would lead to a wide range of "innocents discrimination tests" being implemented by lethal autonomous weapons.

## 2 Review of some Theories

Murphy *et al* [5] advocates for approaches to robot "behaviour" administration, in summary they revolve around concepts such as deployment of the highest legal and professional standards in design, development and use of robots; robots being able to discriminate effective the roles of different human persons and in making such distinction interacting or reacting to them appropriately; robots autonomy should aid its self preservation to the extent that it is functioning properly. There are many innovative concepts in the article such as a black box system for robot perception and operations recording, it is assumed that data obtained would aid investigations if required. Other functions that would be provided for via such data are design improvements, innovation and machine learning.

The protocol put forward in this paper suggests some ways and means that can be used to achieve the tenets that are the *"Alternative Laws of Responsible Robotics" [5],* but the extent to which some of the researchers objectives could be met, may be questionable as they go on to describe robots as being able to attain morality, something that is only in the realm of deep philosophical debates. What would suffice to meet the requirement of the Alternative Laws would be robots to have sufficiently sensitive sensors and processing systems to maximize the benefits of exploitation of sensor data-streams.

When considering machine learning there may be a lot of market hype about artificial intelligence aiding the ultimately programmed robot to achieve great feats, this in turn leads to study and research of a wide range of algorithms. Notably the quality and sensitivity of sensors is the primary determinant of machine learning algorithms, that can be deployed in robotics. One cannot engage in a quest for development of the "ultimate" Deep Learning Algorithms for robots if he or she is not equipped with the "ultimate" sensors, that generate the ultimate data-streams into a robot's control system. To avoid coming up with any unreliable learning methods, this paper suggests where they can be used but does not attempt to produce any specific learning methods.

Text, speech and graphics provide a wide range of sentiment expression – persons may express themselves in relation to various events, processes, practices, etc. these are formally known as domains in the realm of sentiment analysis. An adequately equipped autonomous robot would tap into mass media and other communication/archiving systems with a view to gleaning from the chatter some valuable nuggets of actionable intelligence. Sensors for capturing sentiment are well advanced and of high resolution of primary interest would be their capture and analysis of actual real world events into sentiment analysis reports that are deduced via analytic methods.

Internet chatter, the electromagnetic spectrum based communications, etc. are a vast pool of information on infractions against international humanitarian law in the course of armed conflicts.

**Humanitarian Algorithms : A Codified Key Safety Switch Protocol for Lethal Autonomy [12]**

Sentiment analysis [6] via informatics classifies this communication chatter across appropriate tactical domains which are then consumed as intelligence or humanitarian relief reports.

Deep Learning Algorithms offer new approaches for analyzing layered sensor inputs in form of graphics, speech and text – a participant in subject chatter could be assessed for intellectual depth, personal preferences, and other idiosyncratic traits. But sentiment analysis tools are domain specific yet the fast evolving nature of a conflict may demand that they are equipped with Domain Adaptation capabilities for identifying and classifying chatter from hereto unclassified sources. Deep Learning discovers abstractions from multiple layers of data. In order to achieve Domain Adaptation a source domain classifier must be mapped onto a target domain classifier via an Intermediate Representation.

It would be good to design a protocol and related weapon system with Deep Learning [6] eg. For sentiment analysis, yet this remains an unrealistic expectation, as a result of the extensive demands of such a system specification. Inevitably the unpredictable and fast evolving nature of armed conflicts, requires Deep Learning Algorithms of the Domain Adaptation type but protocols for such systems are beyond the scope of this paper.

Another category of Deep Learning that would contribute towards more efficient autonomous robots is the multi-modal type [7], this is when as system learns of an event by way of different modes eg. Sound and sight, thereafter a multi-modal system can assess and determine any future occurrence purely by either of the modes.

Unsupervised Learning on a lethal autonomous weapon would be acceptable as the basis for decisions, if and only if its results are applied for purposes of constraining and/or inspecting propositions for lethal actions. However should the results of unsupervised learning by lethal autonomous weapons, be used for initiating and executing lethal action, it would imply that any such weapon is self-willed and can make up its "mind" on the best course of action without regard to the tactical objectives of its owners. Nonetheless technology described [7] of multi-modal type deep learning is still a long way from actualization, some of the reasons being:

**Humanitarian Algorithms : A Codified Key Safety Switch Protocol for Lethal Autonomy [12]**

1. Unrealistically long learning times of no utility to immediate tactical mission decision making
2. Computing power described in [7] cannot be hosted on an unmanned aerial vehicle for reasons of its extremely large installation size
3. 15.8% accuracy for object recognition against 20,000 categories is not good enough for real world war-fighting
4. it is not designed for mobile robots.

This Capability vs. Practicality Gap is a certainty when considering a broad spectrum of technologies that seem to have applicability in the domain of lethal autonomy. The implication of this determination is that a technological protocol for controlling lethal autonomy should be founded on practicality and not on seemingly super capabilities that are nowhere near being applied in systems integration.

Arkin [8] offers a technical solution for controlling lethal autonomy, it however lacks a broad range of autonomous capabilities that are difficult to implement, eg.
1. The ethical governor does not incorporate autonomous continuous learning and consultation/collaboration between a lethal autonomous weapon and other related external autonomous platforms
2. It presumes the existence of a "human-adjusting-the-loop"
3. It can only be implemented with a finite set of behavioural patterns scrutiny capability

Veritably it is only possible to make a determination if a lethal autonomous weapon complies with international humanitarian law, if it is captured, dismantled, tested and analyzed by engineers and scientists, eg.
1. its computer program executable code is decompiled
2. its circuit boards, sensors and actuators are analyzed
3. test runs for performance and responses are undertaken

Arkin's description of the Ethical Governor constrains already generated lethal response signals and does not on its own accord generate valid, legal and effective lethal actions. Simply stated in his own

Page 5 of 14

**Humanitarian Algorithms : A Codified Key Safety Switch Protocol for Lethal Autonomy [12]**words the ethical governor is a system bottleneck. With the critical nature of the increased speed of execution of an OODA(Observe, Orient, Decide and Act) Loop, the last thing that one would want on a lethal autonomous weapon would be a system bottleneck. In effect the assumption is that without Arkin's Ethical Governor a lethal weapon would be non-discriminative and disproportional in its responses, hardly what one would expect of a system that exhibits autonomy.

There are more real world problems and situations than are computer programs to handle and react to them. One would easily assume that an insufficient number of constraints are found within the ethical governor, which may in fact turn it into a functionality choke and not a bottleneck. It may be a choke, that prevents a wide range of legitimate lethal actions and in the long run renders an autonomous weapon system that has it installed ineffective for military usage. Another notable problem with the Arkin Ethical Governor is that it does not incorporate any form of autonomous learning within its system design, making it completely dependent on human based adjustment.

**3 The Protocol**

The following steps would be utilized by the suggested safety switch protocols for lethal autonomous weapons in effecting their OODA(Observe, Orient, Decide and Act) Loops:

Step(1) : The lethal autonomous weapon system would activate its target acquisition and designation functions based upon its "world view" and learning capabilities.

Step(2) : A target is identified and its characteristics acquired by the weapon system

Step(3) : Processing of target characteristics is undertaken with a view to labeling the target as neutral, friendly, or hostile. If it determines that the target is hostile an appropriate lethal response is immediately selected

Step(4) : The weapon system is switched on and engages the target

Step(5) : System then conducts a battle damage assessment by way of acquiring new target characteristics to determine if the target has been neutralized

Step(6) : If the target is still hostile Step(3) to Step(5) are conducted again, else it concludes that the target has been neutralized

Step(7) : The Mission is Accomplished or other autonomous lethal weapon systems collaborate to Step(8) : The system learns more about targeting with on-board and off-board data and auto-configuration capabilities

Step(9) : New inspection and constraining methods for lethal action are brought on-line into the system loop during servicing/maintenance

Step(10) : System re-activated and initiates the target searching process

Unlike the Arkin model that claims its system to have the means of behaving morally and determining other parameters such as Ethics and Military Necessity. The safety switch protocol approach assumes that the vital threshold of Military Necessity has been attained and is not a matter to be encoded into the autonomous weapon system – as such the morality, ethics, behaviour, culpability and tactics of the human person who deploys such a system comes into focus. An already fielded weapon system would be in no position to make its determination on issues such as Military Necessity as that would imply that it is in effect self-willed and can make its own military decisions for its own benefit.

Proportionality in the realm of the safety switch protocol type lethal autonomous weapons would imply that an appropriate response is selected during a specific mission in response to a perceived hostile entity. This interpretation would be markedly different from the interpretation of proportionality in response to the action/event by hostiles that occasioned the Military Necessity resulting in the deployment of the autonomous weapon in the first instance. Overall proportionality in the overall prosecution of military operations is again left to human judgment of those who unleash the system into the operational environment as they are best placed to select the appropriate response sets.

<u>Stage 1 – Target Characteristics Acquisition</u>
Fine grained, high resolution type discrimination/distinction of potential targets, is required if the safety switch protocol is to be effected. Target characteristics of interest would be:
1. Place, position and/or location
2. Activity, action, emissions and/or behaviour
3. Possession, use and/or abuse

**Humanitarian Algorithms : A Codified Key Safety Switch Protocol for Lethal Autonomy [12]**

4. Movement and/or direction
5. Grouping, formation, co-ordination, association and/or collaboration
6. Markings, sighting, observations and/or image
7. Sound, noise, voice and/or proclamation, etc.

This list is indeed much longer and dependent upon sensor-types that are available to a lethal autonomous weapon. Setting of pre-mission perceptual thresholds assures one that an autonomous weapon system would not engage targets based on some self-generated unsupervised learning of its operational environment but other forms of unsupervised learning may be permitted eg. In the course of battle damage assessments. This is in effect a notable difference from Arkin's system where once $\rho$-permissible is executed by the autonomous weapon the system cannot learn of its violations but only makes suggestions of potential violations/mistakes to an operator and/or deliberative system before effecting any such action.

Stage 2 – Codified Keys Generation

A safety switch protocol autonomous weapon system utilizes target characteristics from sensor data-streams to generate codified keys that then switch on or constrain the weapon accordingly. A codified key is divided into segments. In this protocol definition no attempt is made to minimize the number of segments in the codified key. For example an autonomous weapon system should have a codified key generated each time it senses a target, even for the same target at different times, different codified keys are generated each with unique time-stamps.

Segmentation of a Codified key takes the following form: eg. Codified key(X), Codified key(Y), etc.
Codified key (X) = {Segment(1) | Segment(2) | Segment (3) | ….. | Segment(n-1) | Segment(n) }

Example of Segment definitions :
Segment(1): Autonomous weapon system identification number
Segment(2): Autonomous weapon system name
Segment(3): Target identification number

# Humanitarian Algorithms : A Codified Key Safety Switch Protocol for Lethal Autonomy [12]

Segment(4): Target identification name

Segment(5): Nature of target : eg. hostile, friend, undetermined or neutral

Segment(6): Time-stamp

Segment(7): Target characteristics from sensor results

Segment(8): Target status eg. Active, dormant, neutralized, undetermined

Segment(9): Target status eg. Combatant or non-combatant

Segment(10): Assumed value of target : e.g. economic value, human life value, strategic value, etc

Segment(11): Resource availability within the weapon system platform, eg. Fuel, endurance, weapons

Segment(n-1): Codified keys of other targets in the vicinity

Segment(n):  Codified keys of associated friendly weapon and support systems in the vicinity

The inclusion of segment terms Segment(n-1) and Segment(n) implies that there could be many other segments within the keys giving the autonomous weapon system ever more functionality.  Codified keys can be exchanged [9: p. 9] between various autonomous weapon systems and command centres in the course of collaboration/co-ordination for purposes of optimization of resources during operations.

The exchange of codified keys between lethal autonomous weapons allows for swapping of operations.  A lethal autonomous weapon should call in other lethal autonomous weapon(s) to conduct an operation for which they have more appropriate mission payloads.  A swap between a mobile and a stationary robot would only require the two robots involved.  In situations where there are more than one mobile target an extra robot is required in most instances to prevent target loss during a swap.

Notably unlike the Arkin approach this protocol does not use only the terminologies combatants and non-combatants since they do not suffice.  In many instances extra information is required because some combatants are friendly, while others are hostile hence the utilization of additional terminologies such as hostile, friendly and neutral in Segments (5), (8) & (9).  Codified keys are critical as they should facilitate :

1. Machine Learning for Situational Awareness and Mission execution by Autonomous weapons
2. Graphic display of the digitized battle-space at a command centre

**Humanitarian Algorithms : A Codified Key Safety Switch Protocol for Lethal Autonomy [12]**

3. Co-operative "thinking" and co-ordination between Autonomous weapons
4. Detailed digital archiving of the evolving battle-space for review by commanders
5. Co-operation between manned and autonomous platforms
6. Black box type functions for investigating weapon system malfunctions and/or errors
7. Research, design, improvements and innovation

With the codified key implementations on lethal autonomous weapons the following types of attacks would be facilitated :

1. Pre-programmed attacks that are executed autonomously
2. Intelligent attacks where the switch learns of and reasons about the target
3. Induced attacks : these are not desired but they must be anticipated. They occur when tampering, deception and/or interference with the switch causes undesired weapons release

A lot of previous work in the domain of lethal autonomy is based on Arkin's research, of great interest is his ability to produce part of the source code of lethal autonomous weapon systems in his published research. But his work is done primarily for the military of the United States of America, the implication being that a lot of it is not published, eg. We have not go the whole range of source code, and hardware on which he bases his work, this paper is an attempt to develop an open, international, and technology neutral approach to handling safety and ethics issues in Lethal Autonomy. Other of his views such as those on classification of autonomy have not found much favour in Kaminski *et al* [10].

Kaminski *et al* [10] details the growing demand for autonomy in military systems in countries such as China and the USA, it also calls for open architectures at least within the military forces of the USA, where any such systems shall be well understood by commanders and operators.

Stage 3 – The Switching System
The Codified keys invoke any one of the switching rules in the third stage. Before the switching rules are invoked a codified key is symmetrically replicated and forwarded to the switching rule modules as well as the command module, if more than one switching rule is invoked the command module shall

# Humanitarian Algorithms : A Codified Key Safety Switch Protocol for Lethal Autonomy [12]

detect the anomaly as a Key Clash as it receives all invoked switch rules and shall wait for another codified key. If action on the basis of the first codified key has not been executed after the key clash, the command module nullifies all invoked actions. Command module also deals with communications.

This safety switch protocol system could be installed on a weapon system platform such as an unmanned aerial vehicle, similarly it can and should also be implemented on specific weapons borne by the unmanned aerial vehicle or robot such as tactical missiles. Sensors continuously pass on data for codified keys into the switching system allowing for implementation of a wide range of reality checks.

Switching Rules

Rule(1): Pre-programmed attack allowed/accepted switch rule – invokes a pre-programmed attack

Rule(2): Not-forbidden switch rule – invites a human-operator to execute the attack

Rule(3): Forbidden switch rule – prevents an attack by the weapon system as it is not permitted by way of Laws of War and/or Rules of Engagement

Rule(4): Attack-aggressor switch rule – autonomous weapon system senses/detects hostile action and reacts by attacking the originating aggressor or their supporting structures

Rule(5): Disarmament switch rule – if mobility fails or capture occurs by way of non-combatants or non-hostiles disarmament occurs and on-board weapons are neutralized

Rule(6): Self-destruct switch rule 1 – if mobility fails destroy the weapon

Rule(7): Self-destruct switch rule 2 – self-destruct if weapon is captured by hostiles

Rule(8): Target-surrendered switch rule – targeted entity indicates by compliance of surrender demand and it therefore tracked continuously but not attacked

Rule(9): "Gotcha" switch rule – an autonomous weapon learns by way of observation, that a target is hostile by way of its characteristics and behaviourial traits and initiates an attack

Rule(10): Mistaken-abort switch rule – eg. A missile flying towards a target having been launched with prior codified keys, deduces that the "man planting a road-side bomb", is actually a toddler playing a game of marbles and automatically aborts the attack

Rule(11): Cessation of hostilities switch rule – effected in conjunction with a command centre eg. robot learns from its time-stamps, that a cease fire time table has come into effect

# Humanitarian Algorithms : A Codified Key Safety Switch Protocol for Lethal Autonomy [12]

Rule(12): Impossible-abort switch rule – switch does not initiate execution of an attack after detecting errors or other faults within the segments of the codified key, ie. A malfunction of the sensors, hardware , software, etc.

Rule(13): By-pass switch rule – codified key noted but does not override previously invoked key

Rule(14): Override switch rule – override already invoked key and switch off action if not initiated or finalized and system must await a new key

An autonomous weapon should have continuous generation and processing of codified keys in relation to a specific target until it is neutralized or the attack is aborted.

**Sketch of Codified Key Safety Switch Protocol System**

--

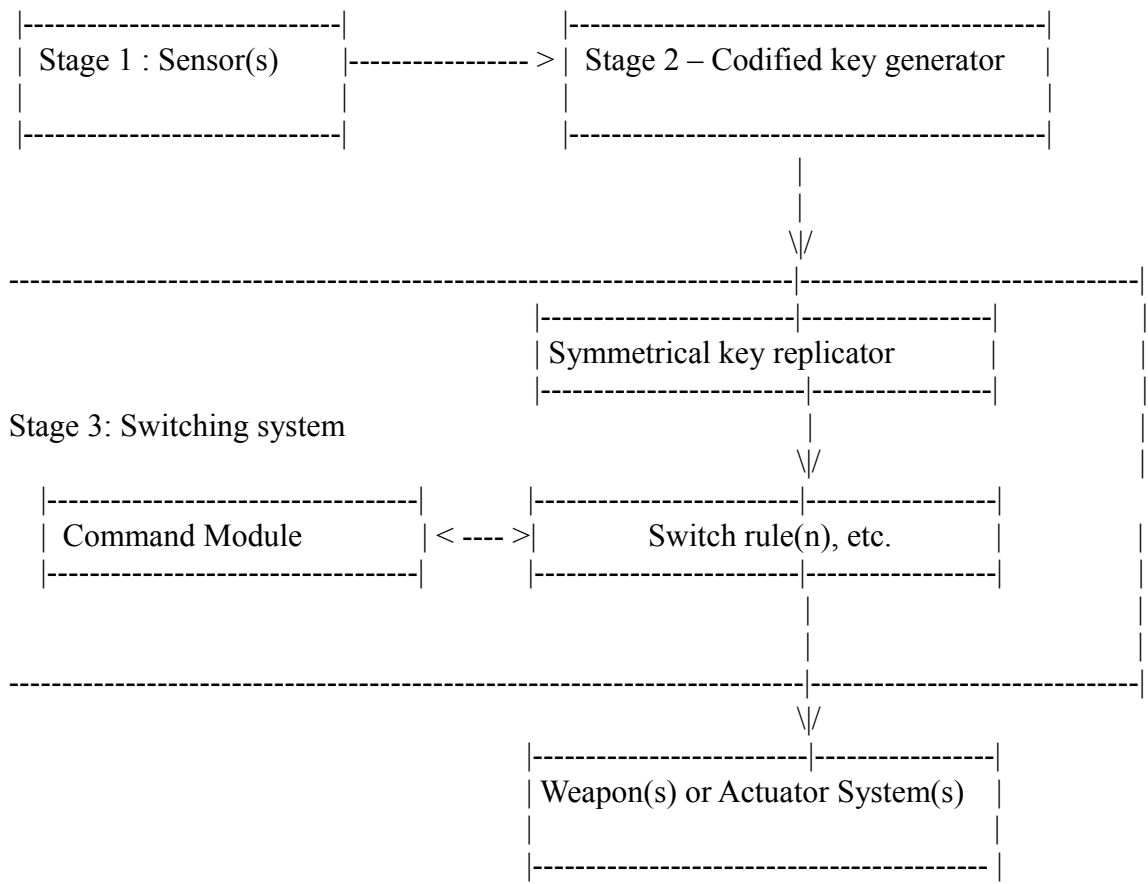

--

**Humanitarian Algorithms : A Codified Key Safety Switch Protocol for Lethal Autonomy [12]**

Fidelity, sensitivity, resolution and high quality of sensor gear on-board a lethal autonomous weapon determines the complexity and sophistication of codified keys generated for safety switching protocols and the details of the functionality that they can initiate and control.

**4 Conclusion**

Inevitably to carry out tasks and to obtain mission objectives, resources are required, this gives a justification for inclusion of Segment(11) in the codified key – the resource availability indicator. Warwela [11] is a detailed theoretical study on the issue of resource management in robots, with a view to attaining operational self-sufficiency.

Implementation of machine learning can occur within the codified key generator and the switching system, but at this stage specific implementations have not been proposed for reasons that they would be specific to any hardware platform and software environment. The implication of incorporating them into a protocol description would be that of developing a proprietary and not open protocol which is not in line with the objectives of this paper.

It is not within the scope of this paper to develop actual computer source code for an autonomous weapon system, but there are some suggestions on how to go about upholding International Humanitarian Law if your are a professional working in the domain of lethal autonomous weapon systems engineering. Hopefully these insights shall lead to a safer, stable and more peaceful world.